\documentclass[twocolumn,english,prl,preview]{revtex4-1}
\usepackage[T1]{fontenc}
\usepackage[latin9]{inputenc}
\usepackage{amsmath}
\usepackage{graphicx}
\usepackage{babel}
\begin{document}
\title{Tailoring the first law of thermodynamics for convective flows}
\author{Karol Makuch}
\email{kmakuch@ichf.edu.pl}

\affiliation{Institute of Physical Chemistry, Polish Academy of Sciences, Kasprzaka
44/52, 01-224 Warsaw, Poland}
\begin{abstract}
Equilibrium thermodynamics is grounded in the law of energy conservation,
with a specific focus on how systems exchange energy with their environment
during transitions between equilibrium states. These transitions are
typically characterized by quantities such as heat absorption and
the work needed to alter the system's volume. This study is inspired
by the potential to develop an analogous, straightforward thermodynamic
description for systems that are out of equilibrium. Here, we explore
the global energy exchanges that occur during transitions between
these nonequilibrium states. We study a system with heat flow and
an external (gravity) field that exhibits macroscopic motion, such
as Rayleigh-Bénard convection. We show that the formula for system's
energy exchange has the same form as in equilibrium. It opens the
possibility of describing out-of-equilibrium systems using a few simple
laws similar to equilibrium thermodynamics.
\end{abstract}
\maketitle

\section{Introduction}

Consider the phenomenon where water, despite being cooled below its
freezing point, remains in a liquid state - an unstable condition
known as supercooling. A minor disturbance, such as a simple shake,
can trigger a rapid transition to solid ice. Equilibrium thermodynamics
provides a comprehensive framework to understand such behaviors, predicting
the stability of equilibrium states by analyzing energy flow and exchanges
with the surrounding environment \citep{Thermodynamics_and_an_Introduction_to_Thermostatistics_2ed_H_Callen}.
But what about systems that are out of equilibrium? Can their stability
also be inferred from energy considerations?

Motivated by this question, this paper explores the possibility of
a thermodynamic-like theory for nonequilibrium systems. Specifically,
we examine systems like a gas undergoing macroscopic heat-driven convection
(illustrated in Fig. 1), which can exhibit abrupt state changes in
response to minor temperature variations \citep{getling1983evolution,eidelman2006hysteresis}.
Developing a predictive theory for these nonequilibrium transitions,
akin to equilibrium thermodynamics, could profoundly enhance our understanding
of atmospheric phenomena and improve designs in areas like steady
state chemical reactors \citep{Elements_of_Chemical_Reaction_Engineering_Fogler__3rd_Edition}.

\begin{figure}
\includegraphics[width=8.5cm]{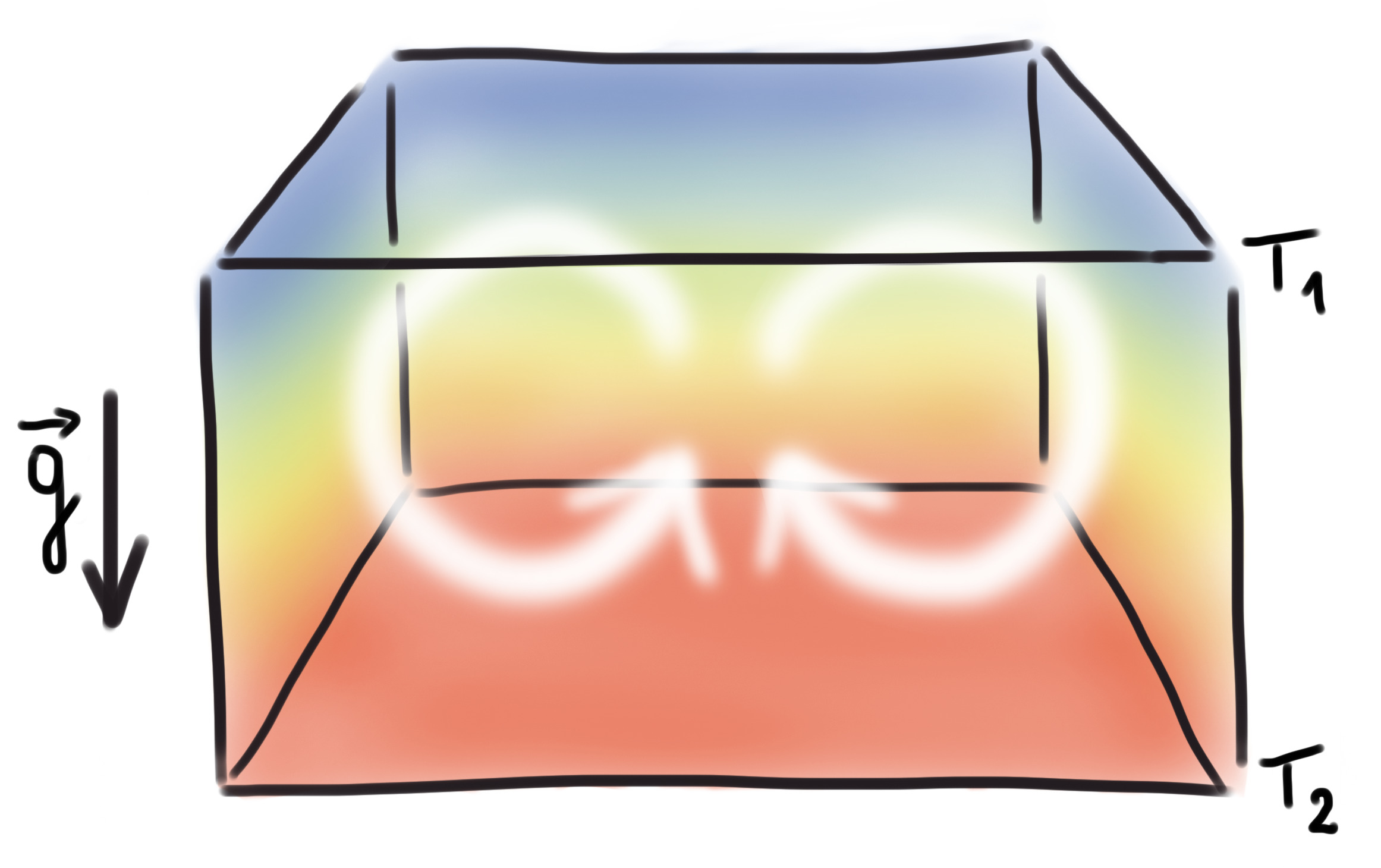}\caption{\label{fig:Ideal-gas-in} Rayleigh-Bénard convection in a fluid. The
figure illustrates the temperature distribution and the resulting
convective flow pattern in a fluid subjected to a temperature gradient.
The bottom boundary, denoted by $T_{2}$, is at a higher temperature,
while the top boundary, denoted by $T_{1}$, is at a lower temperature.
The temperature gradient, combined with the gravitational field ${\bf g}$
acting downward, induces buoyancy-driven convection. Colors schematically
shows temperature profile (red - hotter, blue - colder).}
\end{figure}

To develop a thermodynamic-like theory beyond equilibrium, we concentrated
on a basic aspect of equilibrium thermodynamics: energy balance. This
concept is represented in thermodynamics by the first law. Energy
is a fundamental physical quantity, and the energy balance beyond
equilibrium has been studied in many different contexts. The local
(in infinitesimally small volume) energy balance is the foundation
of linear irreversible thermodynamics \citep{Groot_Mazur_Non-equilibrium_thermodynamics}.
There is a vibrant research field of stochastic thermodynamics related
to energy balance for small systems \citep{sekimoto1998langevin,seifert2012stochastic,van2015ensemble,ding2022unified,manzano2022quantum}.

However, examining the total energy balance in macroscopic nonequilibrium
systems has received limited attention. Even for equilibrium systems,
hydrodynamic-based examination of the total energy balance in transitions
between macroscopic equilibrium states has been performed recently
\citep{swaney2019transport}. Beyond equilibrium, a similar approach
has been used to study of the total energy balance for a quiescent
fluid in a heat flow \citep{holyst2022thermodynamics,holyst2023fundamental,maciolek2023parameters,holyst2023fundamentalGraw}.
These studies show that in a mixture of gases in the presence of heat
flow or a gravity field, the total energy balance has a simple form,
$dE=\mkern3mu\mathchar'26\mkern-12mu dQ+\mkern3mu\mathchar'26\mkern-12mu dW$,
with $\mkern3mu\mathchar'26\mkern-12mu dQ$ being the heat, and $\mkern3mu\mathchar'26\mkern-12mu dW$
is the volumetric work. The total energy balance in these nonequilibrium
systems has the same form as in equilibrium thermodynamics. For more
complicated systems that contain macroscopic flow, such as ideal gas
between parallel plates in a shearing motion that exhibits nonequilibrium
phase transition \citep{makuch2023steady}, the energy balance, $dE=\mkern3mu\mathchar'26\mkern-12mu dQ+\mkern3mu\mathchar'26\mkern-12mu dW+\mkern3mu\mathchar'26\mkern-12mu dW_{w}$,
contains another term $\mkern3mu\mathchar'26\mkern-12mu dW_{w}$ which
does not appear in equilibrium. It represents the excess shear work
of the plate that induces the shearing motion of the gas.

It is worth mentioning that the term \textquotedblleft first law\textquotedblright{}
already appeared in the context of systems with shearing flow \citep{evans1980shear,evans1986rheology}.
However, the term \textquotedblleft first law\textquotedblright{}
used in papers \citep{evans1980shear,evans1986rheology} refers to
the fundamental relation solely for internal energy \citep{daivis2003steady,daivis2008thermodynamic}.
The fundamental relation is another vivid topic \citep{sasa2006steady,nakagawa2019global}
but is beyond the scope of the current paper.

It has been recognized that in steady state systems, there is a component
of the heat exchanged with the environment, in addition to the steady
state heat flow, known as excess heat and denoted as $\mkern3mu\mathchar'26\mkern-12mu dQ$.
This excess heat may play a crucial role in developing a thermodynamic-like
description of steady states \citep{oono1998steady}. The excess heat
by itself has been recently studied by different researchers experimentally
\citep{yamamoto2023calorimetry} and theoretically \citep{boksenbojm2011heat,chiba2016numerical,maes2019nonequilibrium,Dolai2023Towards,dolai2024specific}.

On the one hand, the studies mentioned above offer a promising foundation
for achieving a thermodynamic-like description of nonequilibrium states.
On the other hand, one might be curious that one of the most common
systems has not been approached in a thermodynamic-like manner. Indeed,
heat-driven convection, which serves as a window to understanding
atmospheric phenomena and turbulent systems, still awaits even basic
study. A similar observation has been expressed recently by Yoshimura
and Ito \citep{yoshimura2024two} who claim ``Deterministic hydrodynamic
systems described by the Navier-Stokes equation are among the least
investigated subjects''. Nakagawa and Sasa also made similar observations
discussing the concept of heat in nonequilibrium states \citep{nakagawa2022unique}.

Following this line of reasoning, we derive the global energy balance
for the nonequilibrium system depicted in Fig. \ref{fig:Ideal-gas-in}.
The presented energy balance describes the transitions of the system
due to the change of boundary temperatures, the shape of the system,
and the shift in the gravity field. The resulting global energy balance
mirrors the first law of equilibrium thermodynamics and simplifies
to the equilibrium first law when transitions occur between equilibrium
states. Given that the system in Fig. \ref{fig:Ideal-gas-in} captures
key features of atmospheric dynamics, heat-driven convection, and
turbulence, this achievement represents a significant step toward
formulating a thermodynamic-like description for many out-of-equilibrium
systems.

\section{Balance equations of a single component system with time dependent
potential}

To understand how a nonequilibrium system exchanges energy with its
environment, we assume a hydrodynamic description of the system \citep{Groot_Mazur_Non-equilibrium_thermodynamics}.
The total energy consists of macroscopic kinetic energy (related to
the macroscopic motion determined by the average velocity field $\mathbf{v}$
in the system), $e_{kin}=\mathbf{v}^{2}/2,$ internal, $u$, and gravitational
energy $\varphi$. The total energy density per unit mass is defined
as
\begin{equation}
e=e_{kin}+u+\varphi.\label{eq:total energy-1}
\end{equation}
The flow of the energy is described by corresponding flux
\begin{equation}
\mathbf{J}_{e}=\rho e\mathbf{v}+P\cdot\mathbf{v}+\mathbf{J}_{q},
\end{equation}
which includes convection ($\rho$ is the volumetric mass density)
and fluxes of the energy due to mechanical work, $P\cdot\mathbf{v}$
(the pressure tensor is denoted by $P$), and the heat flux, $\mathbf{J}_{q}$.
The evolution equations in hydrodynamics take the form of a balance
equation,
\begin{equation}
\frac{\partial}{\partial t}f\left(\mathbf{r},t\right)=-\nabla\cdot\mathbf{J}\left(\mathbf{r},t\right)+\sigma\left(\mathbf{r},t\right),\label{eq:gen balance eq-1}
\end{equation}
where $f$ is volumetric density, $\mathbf{J}\left(\mathbf{r},t\right)$
is flux and $\sigma\left(\mathbf{r},t\right)$ is the source term.
We decompose $\mathbf{J}$ into a convective term term, $f\left(\mathbf{r},t\right)\mathbf{v}\left(\mathbf{r},t\right)$,
and the rest (non-convective),
\begin{equation}
\mathbf{J}\left(\mathbf{r},t\right)=f\left(\mathbf{r},t\right)\mathbf{v}\left(\mathbf{r},t\right)+\mathbf{J}_{nc}\left(\mathbf{r},t\right),\label{eq:flux f convection and rest-1}
\end{equation}
which is the definition of non-convective flux $\mathbf{J}_{nc}\left(\mathbf{r},t\right)$.
In particular, with 
\begin{align}
f_{e} & =\rho e,\label{eq:f for total e-1}\\
\mathbf{J}_{nc,e} & =P\cdot\mathbf{v}+\mathbf{J}_{q},\\
\sigma_{e} & =\rho\partial_{t}\varphi,
\end{align}
we obtain the total energy balance equation 
\begin{equation}
\partial_{t}\rho e=-\nabla\cdot\mathbf{J}_{e}+\rho\partial_{t}\varphi.\label{eq:total energy balance with source-1}
\end{equation}
With
\begin{align}
f_{\varphi} & =\rho\varphi,\\
\mathbf{J}_{nc,\varphi} & =0,\\
\sigma_{\varphi} & =\rho\mathbf{v}\cdot\nabla\varphi+\rho\partial_{t}\varphi,
\end{align}
we get the external potential balance equation
\begin{equation}
\partial_{t}\rho\varphi=-\nabla\cdot\left(\rho\varphi\mathbf{v}\right)+\rho\mathbf{v}\cdot\nabla\varphi+\rho\partial_{t}\varphi.\label{eq:ext pot time dependent-1}
\end{equation}
With 
\begin{align}
f_{ekin} & =\rho e_{kin},\\
\mathbf{J}_{nc,ekin} & =P\cdot\mathbf{v},\\
\sigma_{ekin} & =P:\left[\nabla{\bf v}\right]-\rho{\bf v}\cdot\nabla\varphi,
\end{align}
we get the kinetic energy balance equation
\begin{equation}
\partial_{t}\frac{1}{2}\rho{\bf v}^{2}=-\nabla\cdot\left(\frac{1}{2}\rho{\bf v}^{2}{\bf v}+P\cdot{\bf v}\right)+P:\left[\nabla{\bf v}\right]+\rho{\bf v}\cdot\left(-\nabla\varphi\right).\label{eq:kin energy balance pot phi-1}
\end{equation}
With
\begin{align}
f_{u} & =\rho u,\\
\mathbf{J}_{nc,e} & =\mathbf{J}_{q},\\
\sigma_{u} & =-P:\left[\nabla\mathbf{v}\right],\label{eq:sigma for kin e-1}
\end{align}
we obtain the internal energy balance equation
\begin{equation}
\partial_{t}\rho u=-\nabla\cdot\left(\rho u\mathbf{v}+\mathbf{J}_{q}\right)-P:\left[\nabla\mathbf{v}\right].\label{eq:internal energy balance}
\end{equation}
Notice that the sum of the last three cases gives the energy balance
equation (the first case above), because total energy is the sum of
external, kinetic and internal energy. It is worth mentioning that
the above balance equations generalize the considerations presented
in the monograph of Mazur and de Groot \citep{Groot_Mazur_Non-equilibrium_thermodynamics}
to include the time dependent external potential $\varphi\left(\mathbf{r},t\right)$.

We further assume that the above mentioned hydrodynamic densities,
fluxes and sources along with the velocity field are known. To determine
the evolution of energies in practice we would have to supplement
the above description with additional information (equations of states,
constitutive relations, boundary conditions). However, as we will
see, for the investigation of the general structure of the system's
energy exchange with its environment, the above quantities are sufficient
and other parts of hydrodynamic theory are not essential. Energy is
a fundamental physical quantity conserved on the most basic level.
It will be conserved independently of how the heat flux is related
to the temperature field or any other system's properties, such as
those that are considered in rational and extended thermodynamics
\citep{Extended_Irreversible_Thermodynamics_by_David_Jou_Georgy_Lebon}.

\section{Global energy balance equations}

In the previous section we discussed time dependent balance equations
defined at each point in space. These equations describe both the
system and its environment. By the system we mean a part of the space
(a region), confined within the volume $V\left(t\right)$, which may
move over time. For example, in a steady situation presented in Fig.
\ref{fig:Ideal-gas-in}, the gas inside the box has a time independent
volume $V$. However, if we consider the motion of the wall, the region
$V$ will change over time according to the motion of its boundary,
$\partial V\left(t\right)$. The motion of the boundary of the system
defines the velocity of the boundary of the volume, $\mathbf{v}_{V}\left(\mathbf{r},t\right)$.
This velocity is defined only for $\mathbf{r}$ on the boundary of
$V\left(t\right)$.

In the attempt to construct global thermodynamics, we focus on how
the total energy inside the system, $E\left(t\right)\equiv\int_{V\left(t\right)}d^{3}r\,e\left(\mathbf{r},t\right)$,
changes during a process that occurs from an initial time $t_{i}$
to a final time $t_{f}$, $dE\equiv E\left(t_{f}\right)-E\left(t_{i}\right)$.
For any energy represented by its volumetric density $f\left(\mathbf{r},t\right)$,
the total energy $F\left(t\right)$ in the system is given by 
\begin{equation}
F\left(t\right)\equiv\int_{V\left(t\right)}d^{3}r\,f\left(\mathbf{r},t\right),\label{eq:def amount-1}
\end{equation}
and its change after the process by
\begin{equation}
dF\equiv F\left(t_{f}\right)-F\left(t_{i}\right).\label{eq:def delta F tf ti-1}
\end{equation}
From the time derivative of Eq. (\ref{eq:def amount-1}) we get the
rate of change,
\begin{align}
\frac{d}{dt}F\left(t\right) & =\int_{\partial V\left(t\right)}d^{2}r\,\hat{n}\left(\mathbf{r},t\right)\cdot\mathbf{v}_{V}\left(\mathbf{r},t\right)f\left(\mathbf{r},t\right)\nonumber \\
 & +\int_{V\left(t\right)}d^{3}r\,\frac{\partial}{\partial t}f\left(\mathbf{r},t\right),\label{eq:boundary identity}
\end{align}
where $\hat{n}\left(\mathbf{r},t\right)$ is a vector normal to the
surface $\partial V$ pointing outside of the volume \citep{waldmann1967non,swaney2019transport}.
The last term can be expressed in terms of the energy balance Eq.
(\ref{eq:gen balance eq-1}) with the nonconvective energy flux in
Eq. (\ref{eq:flux f convection and rest-1}) giving

\begin{align}
\frac{dF}{dt} & =\int_{\partial V\left(t\right)}d^{2}r\,\hat{n}\left(\mathbf{r},t\right)\cdot\left[\mathbf{v}_{V}\left(\mathbf{r},t\right)-\mathbf{v}\left(\mathbf{r},t\right)\right]f\left(\mathbf{r},t\right)\nonumber \\
 & -\int_{\partial V\left(t\right)}d^{2}r\,\hat{n}\left(\mathbf{r},t\right)\cdot\mathbf{J}_{nc}\left(\mathbf{r},t\right)+\int_{V\left(t\right)}\,\sigma\left(\mathbf{r},t\right).\label{eq:rate F full}
\end{align}
The above formula for the change of energy in the system (understood
as a region in space) uses a small number of assumptions. In the above
derivation we used Gauss's theorem, $\int_{V\left(t\right)}d^{3}r\,\nabla\cdot\mathbf{J}_{nc}\left(\mathbf{r},t\right)=\int_{\partial V\left(t\right)}d^{2}r\,\hat{n}\left(\mathbf{r},t\right)\cdot\mathbf{J}_{nc}\left(\mathbf{r},t\right)$.
Essentially, up to this point, we assumed that the energy densities
are given by hydrodynamic balance equations and calculated their changes
in a process. From now on, we assume that the system is closed, so
there is no flow through the surface $\partial V$. This means that
the flux of the particles through the surface $\partial V$ vanishes
in the reference frame where the surface element is at rest,
\begin{equation}
\hat{n}\left(\mathbf{r},t\right)\cdot\left[\mathbf{v}_{V}\left(\mathbf{r},t\right)-\mathbf{v}\left(\mathbf{r},t\right)\right]=0\text{ for }\mathbf{r}\in\partial V.\label{eq:no flux through surface-1}
\end{equation}
As before, $\hat{n}\left(\mathbf{r}\right)$ is the vector normal
to the surface, pointing outside the region $V$. The above condition
simplifies the rate of energy change (\ref{eq:rate F full}) to
\begin{equation}
\frac{dF}{dt}=-\int_{\partial V\left(t\right)}d^{2}r\,\hat{n}\left(\mathbf{r},t\right)\cdot\mathbf{J}_{nc}\left(\mathbf{r},t\right)+\int_{V\left(t\right)}d^{3}r\,\sigma\left(\mathbf{r},t\right).\label{eq:rate F closed}
\end{equation}
By integrating the rate of change over time we get the change of energy,
$dF\equiv F\left(t_{f}\right)-F\left(t_{i}\right)=\int_{t_{i}}^{t_{f}}dF\left(t\right)/dt$,
in the following form, 

\begin{align}
dF= & -\int_{t_{i}}^{t_{f}}dt\int_{\partial V\left(t\right)}d^{2}r\,\hat{n}\left(\mathbf{r},t\right)\cdot\mathbf{J}_{nc}\left(\mathbf{r},t\right)\nonumber \\
 & +\int_{t_{i}}^{t_{f}}dt\int_{V\left(t\right)}d^{3}r\,\sigma\left(\mathbf{r},t\right).\label{eq:change F closed}
\end{align}

The above two equations describe the energy change of a closed hydrodynamic
system. We use the former equation to study steady state and the latter
equation to study energy exchange during transitions between steady
states. In further analysis, we assume slip boundary conditions on
the boundaries of a closed system. This means that on the wall, the
transverse components of the pressure tensor vanish,
\begin{equation}
\hat{n}\left(\mathbf{r}\right)\cdot P\left(\mathbf{r}\right)\cdot\left(\mathbf{1}-\hat{n}\left(\mathbf{r}\right)\hat{n}\left(\mathbf{r}\right)\right)=0\text{ for }\mathbf{r}\in\partial V.\label{eq:slip P vanishing-1}
\end{equation}
The no-slip boundary condition can also be used without changing the
main result of the paper.

The above assumptions are sufficient to generalize the first law of
thermodynamics to nonequilibrium closed systems including the system
shown in Fig. \ref{fig:Ideal-gas-in}. The application of Eq. (\ref{eq:rate F closed})
to the case of external, kinetic, internal and total energy (cf. Eqs.
(\ref{eq:f for total e-1}-\ref{eq:sigma for kin e-1})) yields the
following:
\begin{equation}
\frac{d\Phi}{dt}=\int_{V\left(t\right)}d^{3}r\,\rho\mathbf{v}\cdot\nabla\varphi+\int_{V\left(t\right)}d^{3}r\,\rho\partial_{t}\varphi,\label{eq:rate Phi-1}
\end{equation}
\begin{align}
\frac{dE_{kin}}{dt} & =-\int_{\partial V\left(t\right)}d^{2}r\,\hat{n}\left(\mathbf{r},t\right)\cdot P\cdot\mathbf{v}+\int_{V\left(t\right)}d^{3}r\,P:\left[\nabla{\bf v}\right]\nonumber \\
 & -\int_{V\left(t\right)}d^{3}r\rho{\bf v}\cdot\nabla\varphi,\label{eq:rate Ekin-1}
\end{align}
\begin{equation}
\frac{dU}{dt}=-\int_{\partial V\left(t\right)}d^{2}r\,\hat{n}\left(\mathbf{r},t\right)\cdot\mathbf{J}_{q}\left(\mathbf{r},t\right)-\int_{V\left(t\right)}d^{3}r\,P:\left[\nabla\mathbf{v}\right],\label{eq:rate U-1}
\end{equation}
\begin{align}
\frac{dE}{dt}= & -\int_{\partial V\left(t\right)}d^{2}r\,\hat{n}\left(\mathbf{r},t\right)\cdot\mathbf{J}_{q}\left(\mathbf{r},t\right)\nonumber \\
 & -\int_{\partial V\left(t\right)}d^{2}r\,\hat{n}\left(\mathbf{r},t\right)\cdot P\cdot\mathbf{v}+\int_{V\left(t\right)}d^{3}r\,\rho\partial_{t}\varphi.\label{eq:rate E-1}
\end{align}
All of the above rates (left hand side) vanish in a steady state,
and the above equations give essential information on the energy balance
at a given steady state.

Let's start with Eq. (\ref{eq:rate Phi-1}) for external potential
energy. At a steady state the external potential does not change,
$\partial_{t}\varphi=0$. Therefore, Eq. (\ref{eq:rate Phi-1}) yields, 

\begin{equation}
\int_{V\left(t\right)}d^{3}r\,\rho\mathbf{v}^{st}\cdot\nabla\varphi=0.\label{eq:only net grav work-1}
\end{equation}
It follows that the gravitational field globally does not perform
work in a steady state because for a homogeneous gravity field pointing
downward in the vertical $z$-coordinate, $\nabla\varphi=\hat{e}_{z}g$
holds, and Eq. (\ref{eq:only net grav work-1}) results in the vanishing
of the center of mass velocity, $\int_{V\left(t\right)}d^{3}r\,\rho\mathbf{v}^{st}=0.$
Therefore, in the reference frame where the boundaries of the system
are immobile, the center of mass velocity vanishes, the position of
the center of mass is constant and gravity does not perform work.
This conclusion also holds for any potential field, like the gravity
field around the earth, in a system with the mass continuity equation
$\partial_{t}\rho=-\nabla\left(\rho\mathbf{v}\right)$. 

Eq. (\ref{eq:only net grav work-1}) simplifies the steady state form
of Eq. (\ref{eq:rate Ekin-1}). Moreover, the system is closed, so
on the surface we have $\hat{n}\left(\mathbf{r},t\right)\cdot\mathbf{v}^{st}=0,$
and the application of the slip boundary conditions given by Eq. (\ref{eq:slip P vanishing-1})
yields,
\begin{equation}
\int_{\partial V\left(t\right)}d^{2}r\,\hat{n}\left(\mathbf{r},t\right)\cdot P^{st}\cdot\mathbf{v}^{st}=0.\label{eq:onle net pres tens work-1}
\end{equation}
The last two equations applied to Eq. (\ref{eq:rate Ekin-1}) at steady
state, where $dE_{kin}/dt=0$, yield

\begin{equation}
\int_{V\left(t\right)}d^{3}r\,P^{st}:\left[\nabla\mathbf{v}^{st}\right]=0.\label{eq:only net diss-1}
\end{equation}
The above term appears locally in hydrodynamic equations for kinetic
and internal energy balance (\ref{eq:kin energy balance pot phi-1}),
(\ref{eq:internal energy balance}). It describes the work of mechanical
forces inside the fluid, typically appearing during expansion and
viscous dissipation. Eq. (\ref{eq:only net diss-1}) indicates that
although the expansion-dissipation exists locally in a steady state,
it globally disappears. 

Finally, the vanishing of the global mechanical work inside the fluid
given by Eq. (\ref{eq:only net diss-1}) simplifies the steady state
form of Eq. (\ref{eq:rate U-1}), which in steady state becomes
\begin{equation}
\int_{\partial V\left(t\right)}d^{2}r\,\hat{n}\left(\mathbf{r},t\right)\cdot\mathbf{J}_{q}^{st}\left(\mathbf{r},t\right)=0.\label{eq:only net heat-1}
\end{equation}
This implies that although heat may locally enter or leave the system
in a steady state, the total heat does not flow into the system. Consequently,
there is no need to introduce the notion of 'excess heat' \citep{oono1998steady},
which is the heat in addition to the constant steady state heat flux.

It is worth noting that Eqs. (\ref{eq:only net grav work-1}-\ref{eq:only net heat-1})
also hold for quasi-steady states, i.e. when there exists a time scale
for which the system can be effectively treated as being in a steady
state. In this case we would consider time averages of Eqs. (\ref{eq:rate Phi-1}-\ref{eq:rate E-1}).
With a similar reasoning we obtain Eqs. (\ref{eq:only net grav work-1}),
(\ref{eq:onle net pres tens work-1}) and (\ref{eq:only net heat-1})
with the symbol $'st'$ denoting the average over a sufficiently long
time scale. Eq. (\ref{eq:only net diss-1}) would change to $\text{\ensuremath{\int_{V\left(t\right)}\ensuremath{d^{3}r\,\left\{ P:\left[\nabla\mathbf{v}\right]\right\} ^{st}}=0}}$
, with $\left\{ P:\left[\nabla\mathbf{v}\right]\right\} ^{st}$ denoting
the time average of $P:\left[\nabla\mathbf{v}\right]$ in a given
quasi-steady state.

The above equations apply to a steady state with convection that is
schematically shown in Fig. \ref{fig:Ideal-gas-in} and also to other
steady states including e.g. those with much more complicated patterns
of temperature and velocity fields \citep{lohse2024ultimate}. These
steady state equations give important insight into the rate of the
global energy exchange of such systems with their surroundings. From
the perspective of the global energy of the system, gravity does not
perform work, there is no global compression-dissipation, and there
is no heat. Thus, from the perspective of global energy, the system
appears to be in equilibrium (no global flux of heat, no global work).

In what follows we study the energy exchange in transitions between
steady states. We assume that at the initial time $t_{i}$ the system
is at a steady state, which is then disturbed by a small change of
boundary temperatures, external gravitational field or the motion
of the surrounding wall. After time $t_{f}$ the system reaches another
steady state. The change in energy is described in this situation
by Eq. (\ref{eq:change F closed}), which applied to the cases of
external, kinetic, internal and total energy (cf. Eqs. (\ref{eq:f for total e-1}-\ref{eq:sigma for kin e-1}))
yields,
\begin{equation}
d\Phi=-\mkern3mu\mathchar'26\mkern-12mu dW_{\varphi}+\mkern3mu\mathchar'26\mkern-12mu dW_{dt\varphi},\label{eq:phi balance net-1}
\end{equation}
\begin{equation}
dE_{kin}=\mkern3mu\mathchar'26\mkern-12mu dM_{S}-\mkern3mu\mathchar'26\mkern-12mu dP_{V}+\mkern3mu\mathchar'26\mkern-12mu dW_{\varphi},
\end{equation}
\begin{equation}
dU=\mkern3mu\mathchar'26\mkern-12mu dQ+\mkern3mu\mathchar'26\mkern-12mu dP_{V},
\end{equation}
\begin{equation}
dE=\mkern3mu\mathchar'26\mkern-12mu dQ+\mkern3mu\mathchar'26\mkern-12mu dM_{S}+\mkern3mu\mathchar'26\mkern-12mu dW_{dt\varphi},\label{eq:E balance net-1}
\end{equation}
where we used the following definitions of the heat differential,
\begin{equation}
\mkern3mu\mathchar'26\mkern-12mu dQ=-\int_{t_{i}}^{t_{f}}dt\int_{\partial V}d^{2}r\,\hat{n}\cdot\mathbf{J}_{q}\left(t\right),\label{eq:def net heat-1}
\end{equation}
volumetric mechanical work differential,
\begin{equation}
\mkern3mu\mathchar'26\mkern-12mu dP_{V}=-\int_{t_{i}}^{t_{f}}dt\int_{V\left(t\right)}d^{3}r\,P:\left[\nabla{\bf v}\right],\label{eq:def diff pv-1}
\end{equation}
mechanical surface force differential,

\begin{equation}
\mkern3mu\mathchar'26\mkern-12mu dM_{S}=-\int_{t_{i}}^{t_{f}}dt\int_{\partial V\left(t\right)}d^{2}r\,\hat{n}\cdot P\cdot{\bf v},\label{eq:def diff ms-1}
\end{equation}
potential work differential,
\begin{equation}
\mkern3mu\mathchar'26\mkern-12mu dW_{\varphi}=-\int_{t_{i}}^{t_{f}}dt\int_{V\left(t\right)}d^{3}r\rho{\bf v}\cdot\nabla\varphi,\label{eq:def diff wphi-1}
\end{equation}
and potential source differential,
\begin{equation}
\mkern3mu\mathchar'26\mkern-12mu dW_{dt\varphi}=\int_{t_{i}}^{t_{f}}dt\int_{V\left(t\right)}d^{3}r\,\rho\partial_{t}\varphi.\label{eq:def diff change time phi-1}
\end{equation}
Formula (\ref{eq:E balance net-1}) is the main result of the paper.
It expresses the balance of total energy and is a generalization of
the first law of equilibrium thermodynamics. The total energy changes
due to heat, work on the surface of the system, and the work of the
external potential field. Without the gravity field ($\mkern3mu\mathchar'26\mkern-12mu dW_{dt\varphi}=0$)
and neglecting macroscopic motion ($dE=dU$), the above formula reduces
to the equilibrium first law, $dU=\mkern3mu\mathchar'26\mkern-12mu dQ-pdV$,
where we use the fact that under these conditions we get $\mkern3mu\mathchar'26\mkern-12mu dM_{S}=-pdV$
\citep{holyst2022thermodynamics}.

\section{Summary}

This study is driven by the possibility of extending thermodynamic
principles to describe systems that are not in equilibrium. As a preliminary
step, we examine total energy exchange in a closed system with macroscopic
motion, exemplified by Rayleigh-Bénard convection. Consequently, we
extend the first law of equilibrium thermodynamics beyond equilibrium
to the form $dE=\mkern3mu\mathchar'26\mkern-12mu dQ+\mkern3mu\mathchar'26\mkern-12mu dM_{S}+\mkern3mu\mathchar'26\mkern-12mu dW_{dt\varphi}$.
This is valid for transitions between steady states and for turbulent
states that exhibit steady-state-like behavior over time. For transitions
between equilibrium states, this generalization reduces to the equilibrium
first law. 

Two aspects are particularly noteworthy. First, the nonequilibrium
first law that we derive retains the same form as the equilibrium
first law. The heat and work differentials are physically interpretable
in a manner akin to equilibrium thermodynamics. Second, in equilibrium
states, viscous dissipation, compression, heat, and mechanical work
vanish locally by definition. In nonequilibrium steady states, although
heat, surface mechanical work, and viscous dissipation-compression
may exist locally, they vanish globally.

This similarity enables a global thermodynamic description of steady
states, comparable to equilibrium systems, and paves the way for studying
heat and surface work differentials. Heat should be measured by monitoring
the total heat on the system's surface, and work differentials by
measuring the force on the wall and its displacement. This approach
opens avenues for numerical and experimental investigations of out-of-equilibrium
systems. These include systems relevant to climate dynamics and steady
state chemical reactors in industry. For instance, understanding heat-driven
convection is fundamental in meteorology, where it plays a critical
role in weather patterns and climate dynamics. In industrial applications,
heat-driven convection is essential in designing more efficient cooling
systems, chemical reactors, and energy generation processes.

Importantly, the heat in transitions between steady states has never
been studied for systems with heat-driven convection. This reveals
an entirely unexplored field of global thermodynamics, currently in
its infancy. From the perspective of a potentially existing thermodynamic
description of nonequilibrium states, we might be at a similar juncture
to where thermodynamics was when Carnot and Clausius began studying
heat differentials.
\begin{acknowledgments}
I thank Robert Ho\l yst and Pawe\l{} \.{Z}uk for engaging in thought-provoking
conversations. Thanks to Natalia Pacocha for her work in creating
the graphics. 
\end{acknowledgments}

\bibliographystyle{plain}

\end{document}